\documentclass[twocolumn,aps,prl,showpacs, superscriptaddress, amsmath,amssymb]{revtex4}

\usepackage{graphicx}
\usepackage{dcolumn}
\usepackage{bm}

\begin{document}

\title{Packing structure of a two-dimensional granular system through the jamming transition}
\author{Xiang Cheng}
\affiliation{The James Franck Institute and Department of Physics,
The University of Chicago, Chicago, Illinois 60637, USA}
\affiliation{Department of Physics, Cornell University, Ithaca,
New York 14853, USA}
\email{xc92@cornell.edu}

\date{\today}

\begin{abstract}

 We have performed a novel experiment on granular packs composed of automatically swelling particles. By analyzing the Voronoi structure of packs going through the jamming transition, we show that the local configuration of a jamming pack is strikingly similar to that of a glass-forming liquid, both in terms of their universal area distribution and the process of defect annealing. Furthermore, we demonstrate that an unambiguous structural signature of the jamming transition can be obtained from the pair correlation functions of a pack. Our study provides insights into the structural properties of general jamming systems.

\end{abstract}

\maketitle

Nearly 300 years ago, Stephen Hales, a highly gifted English scientist and clergyman, designed a clever experiment to investigate the packing structure of spherical particles \cite{Aste}. He put spherical peas into a closed jar filled with water. Absorbing the water slowly, the peas confined in the jar expanded and deformed under the enormous stress of swelling. From the shape of deformed peas, Hales deduced the structure of the initial pea pack. Though this was probably not Hales' intention, his classic experiment provides an excellent way to probe the jamming transition of a granular pack. In this paper, we report a two dimensional (2D) version of Stephen Hales' experiment. From the position of swelling particles in a cell of fixed area, we investigated the structure of a granular pack as it evolves continuously through the jamming transition.

The jamming transition and the jamming phase diagram are intriguing concepts for unifying different transitions in various systems ranging from granular matter to molecular glasses \cite{Liu,O'Hern,Silbert2,Ellenbroek,Olsson,Henkes,Wyart2}. Many interesting features have been found in athermal granular systems near the jamming transition \cite{O'Hern,Silbert2,Ellenbroek,Olsson,Henkes,Wyart2}. However, most of these results were obtained only in theories and simulations for ideal systems of frictionless spherical particles. It is not straightforward how such results should be extended to experimental granular systems with frictional contacts. Majmudar {\it et al.} found that the increasing of both the average coordination number of particles and the pressure of the system agree with the mean-field theory of frictionless particles as the jamming transition is approached \cite{Majmudar}. However, other measurements produce different results in experiments and simulations \cite{Jacob,Bonneau,Bandi}. For example, experiments showed that the ratio of the shear modulus to the bulk modulus of a pack, $G/B$, stays constant as the pressure of the system approaches zero \cite{Jacob,Bonneau}, which contradicts the results of simulations and theories with frictionless particles where $G/B$ diminishes with the pressure \cite{O'Hern,Wyart2}. Moreover, very few experiments have directly investigated the structural signature of the jamming transition. Corwin {\it et al.} indirectly deduced the structural signature of the transition from the statistics of the contact forces between particles \cite{Corwin}. Other than granular systems, Zhang {\it et al.} investigated the structural signature of a colloidal system at finite temperature \cite{Zhang}.

Here, we experimentally study the structure of a granular pack going
through the jamming transition. Instead of peas, we used tapioca
pearls as our granular material. Tapioca pearls are spherical particles
made of starch that swell up by absorbing water.  The
average diameter of dry pearls is $d_0=3.3$ mm with the
polydispersity of $8\%$, which prevents the system from forming
large crystalline areas. A fully swelled particle can be as large as
1.7 times of its original size, which leads to almost 3 times
increases in packing fraction ($\phi$) of a 2D system. More
importantly, the swelling of pearls is slow and uniform, and the pearls maintain their spherical shape during swelling. The polydispersity of swelled particles becomes slightly smaller than that of dry particles.

\begin{figure}
\begin{center}
\includegraphics[width=3.35in]{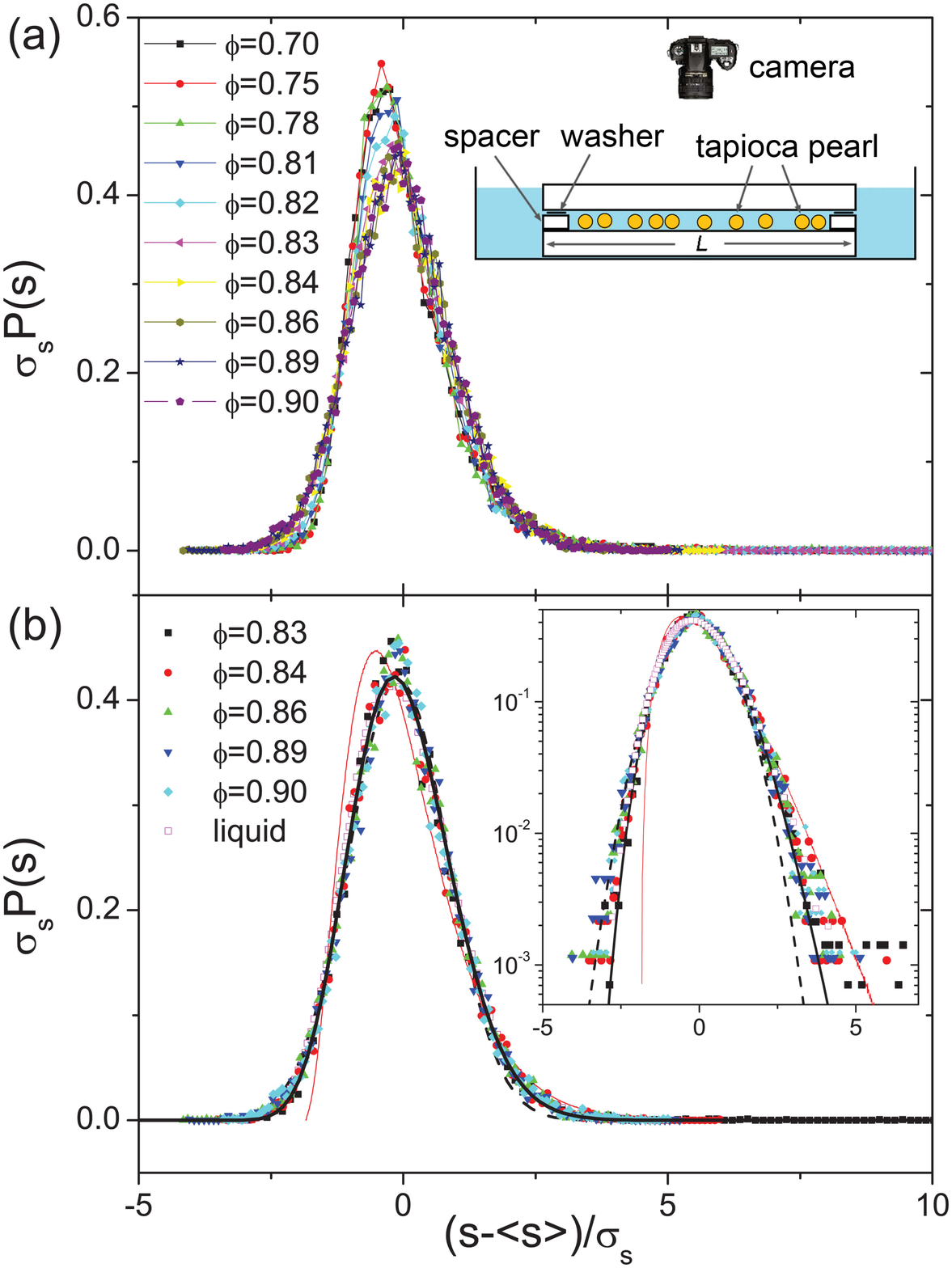}
\end{center}
\caption{Distribution of Voronoi cell areas, shifted
by the average area and scaled by its standard deviation. (a)
Distribution for a pack with $\phi$ increasing from 0.70 to 0.90. Upper right inset shows a schematic of the experimental setup. The spacer and washers keep the particles in one layer. The washers allow water to flow in and out of the otherwise closed cell.
(b) The
collapse of the distributions near and after the jamming transition.
The universal
distribution of Voronoi structure of the glass-forming liquids
(empty magenta squares) is taken from \cite{Starr}. The thick black line is the gamma distribution with $k=30$ and $\theta=0.175$. The black dashed line is a Gaussian fit. The thin
red line is the distribution of Voronoi structure of random points.
The inset is the log-linear plot of the same data.} \label{Figure1}
\end{figure}

For each experiment, a single layer of pearls
is laid down randomly into a square cell with the side length of $L=54.6$ cm.
The vertical height of the cell is kept at 5.454 mm using a spacer and washers, which prevents
{\it fully} swelled pearls from buckling into two layers (Fig.~\ref{Figure1}a inset). Since the jamming transition occurs before pearls fully swell, they are not stuck between the two confining plates near the jamming point. The cell can
hold over 15,000 particles at the initial packing fraction $\phi_0
\gtrsim 0.63$. A subset of the sample (about 10,000 particles) is
studied in the central area of the cell to avoid any
boundary effects. To allow the pearls to swell, the whole system
including the cell and particles is submerged under water. Static friction between particles and the bottom plate is also reduced due to the lubrication of water. A high resolution camera is used to take an image of the sample every 20 s.
Due to the slow swelling of pearls, the system is quasi-stationary
at this time scale. The acquired images are then analyzed with a
particle tracking algorithm \cite{Crocker}. The center of particles
can be determined with an error of 0.1 mm (about $3\%$ of a particle
diameter). A 2D projection of images is used to obtain $\phi$. More details of the experiment can be found in
\cite{Cheng}.

We initially prepared the sample in a dense but unjammed state.  As
the size of particles increases, the system goes through the jamming
transition and is eventually trapped inside the jammed phase. We
constructed a Voronoi tessellation from the centers of particles for
each image with increasing $\phi$. The Voronoi cell of a particle is the polygon consisting of all points closer to the center of the particle than to
centers of any other particles, which reveals local
configurations of packings \cite{Aste}.

We first studied the area distribution of the Voronoi cells,
$P(s)$, at different $\phi$ as the system goes through the jamming
transition. We use a scaling that has been previously used in analyzing the structure of glass-forming liquids \cite{Starr}. The area of each cell $s$ is shifted by the
average area of all cells $\langle s\rangle $ and then is normalized by the standard
deviation $\sigma_s$, {\it i.e.} $\bar{s}=(s-\langle s \rangle)/\sigma_s$.
The probability distribution function is then $P(\bar{s})=\sigma_s
\cdot P(s)$ (Fig.~\ref{Figure1}). The shape of normalized
distributions at small $\phi$ depends on the initial random configuration of
particles (Fig.~\ref{Figure1}a), which varies for
each run. However, as $\phi$ increases toward the jamming point, we
found that all $P(\bar{s})$  collapse into a single master curve. For clarity, the collapsed $P(\bar{s})$ is plotted separately in Fig.~\ref{Figure1}b and also in a log-linear plot in the inset of Fig.~\ref{Figure1}b. It is surprising that all the packs
with $\phi$ ranging continuously from $0.83$ up to $0.90$ (which is
the upper limit of our experiment due to the finite swelling
potential of particles under compression) follow the same universal
distribution. From the structural signature of the jamming transition
which we shall show later, we know that the system jams at
$\phi^*\simeq0.84$. Hence, the universal distribution emerges only close to
and after the jamming transition. It should be emphasized that the
universal distribution for this large range
of $\phi$ is non-trivial: the jammed phase of a pack is far from
stationary. Though particles are confined by their
neighbors, the shape of Voronoi cells is highly mobile as illustrated below. Some
localized collective rearrangement of particles also exists in the
jammed phase \cite{Cheng}.

Quantitatively, we fitted the universal distribution with a shifted gamma
distribution (Fig.~\ref{Figure1}b),
\begin{equation}
\label{gammadistriubiton} {P(\bar{s})=
(\bar{s}-\bar{s}_0)^{k-1}\frac{\exp(-(\bar{s}-\bar{s}_0)/\theta)}{\Gamma(k)\theta^k}},
\end{equation}
where $\theta$ and $k$ are standard parameters of gamma distribution
and $\bar{s}_0$ is a fitting parameter due to the shift in the $s$ axis.
The same distribution has been suggested before in
\cite{Aste2,Kumar}. We also compare the gamma distribution with a
Gaussian distribution (Fig.~\ref{Figure1}b), and find that due to the asymmetric shape of the
distribution, the gamma distribution gives a better description of
our data.

Disorder and geometric constraint are two salient features of the jamming transition and the basic organizing themes underlying the jamming phase diagram. Both of them are essential in determining the shape of the universal distribution, which can be seen as following. First, it is evident that for an ordered structure the area distribution of Voronoi cells has zero variance and therefore should be a delta function (or delta functions). Second, we compare the distribution of our jamming system with that of totally random points without any geometric constraint due to particle size (Fig.~\ref{Figure1}b). For random points, the peak of the distribution shifts to the left and the distribution has longer tail toward the right.

It is known that the Voronoi structure of glass-forming liquids also follows a universal distribution at different temperatures ($T$) ranging from liquid states to supercooled states \cite{Starr}. Even more striking, the universal distribution of the Voronoi structure of glass-forming liquids is {\it quantitatively} similar to the universal distribution of our 2D jamming granular system (Fig.~\ref{Figure1}b). It is likely that near the glass transition the interaction between molecules of a supercooled liquid is dominated by its strong repulsive potential, therefore, like the athermal granular system, the geometric constraint of the molecules determines the local configurations and its distribution. In addition, it is noted that deep inside the jammed phase the area distribution is invariant, which is consistent with previous study of the Voronoi structure of static granular packs at several discrete $\phi$ \cite{Aste2}. Here, we show the invariant distribution with $\phi$ varying continuously and uniformly in a larger range. The universal distributions of Voronoi structure in glass-forming liquids at different $T$ and in a granular system going through the jamming transition at different $\phi$ corroborate the concept of the jamming phase diagram \cite{Liu}.

\begin{figure}
\begin{center}
\includegraphics[width=3.35in]{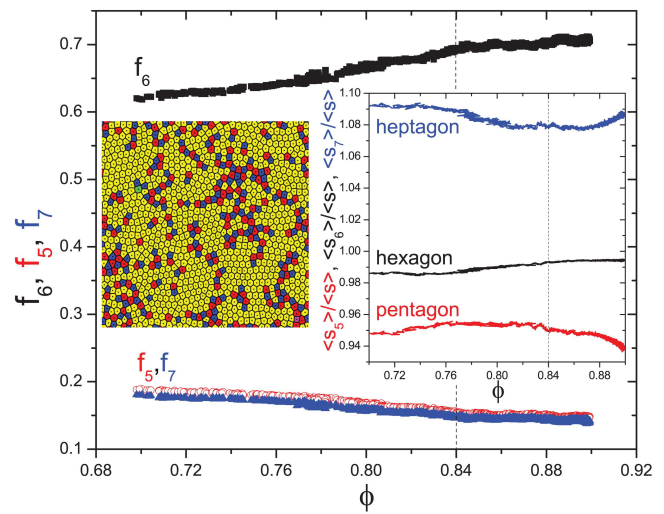}
\end{center}
\caption{
Main plot: Number fraction of different type of cells with increasing $\phi$. Black squares for hexagons
($f_6$), red circles for pentagons ($f_5$) and blue triangles
for heptagons ($f_7$). Left inset: Shape of cells in Voronoi
tessellation of a pack at $\phi=0.86$. Yellow cells are hexagons;
blue cells are pentagons and red cells are heptagons. There is also
one octagon (green) and one square (magenta). Note that most of
pentagons and heptagons join into neighboring pairs and can anneal to hexagons later. Right inset: the
average area of hexagons, pentagons and heptagons over the total
average area of cells.
The vertical dashed lines in both plots indicate the jamming point
obtained from the structural signature of the jamming transition explained in the text.} \label{Figure2}
\end{figure}

We also investigate the shape of Voronoi cells. In 2D, the geometric
Euler constraint implies that average number of edges of a cell
should be 6 \cite{Aste,Perera}. Cells with the number of edges
other than 6 are usually referred to as ``defects''
\cite{Perera,Aharonov,Hentschel}. We count the fractions of
pentagons, hexagons and heptagons in the system ($f_5$,$f_6$ and
$f_7$) as it goes through the jamming transition. As shown in Fig.~\ref{Figure2}, $f_6$
increases while $f_5$ and $f_7$ decrease: with increasing $\phi$, a pentagon-heptagon pair can anneal to two hexagons. It is
interesting that the annealing process continues even at $\phi$
clearly above the jamming point. The shape of the Voronoi cells is
highly mobile due to tiny deformations of particles inside the
jammed phase. Nevertheless, the annealing slows down deep inside the
jammed phase and eventually stops. Due to the disordered nature of the
jammed phase, a finite amount of the pentagon-heptagon pairs
persist \cite{Perera}.

It has been proposed that the glass transition of binary-mixture
molecular liquids can be characterized by the disappearance of
certain kinds of ``liquid-like'' defects (heptagons enclosing small molecules and
pentagons enclosing large molecules) in Voronoi tessellation
\cite{Aharonov,Hentschel}. It is possible that in our granular
system the ``liquid-like'' defects diminish during annealing and the remaining defects inside jammed phase are
associated with ``glass-like'' defects. Due to the continuous
distribution of the size of our particles and the difficulty to
measure the accurate size of particles under water, we cannot
directly verify the above scenario. However, a consistency check
can be made by analyzing the average area of different types of
cells. If heptagons enclosing small molecules, which should have
smaller Voronoi cells in a condensed phase on average, diminish,
then the average area of heptagons normalized by the total average
area, should increase. Similarly, if pentagons enclosing
large molecules disappear, then the average area of pentagons over the
total average area should decrease. We measured the normalized average area of
pentagons, hexagons and heptagons, $\langle s_{5} \rangle/\langle s
\rangle$, $\langle s_{6} \rangle/\langle s
\rangle$ and $\langle s_{7} \rangle/\langle s
\rangle$, as a function of $\phi$ in our experiments (Fig.~\ref{Figure2} right inset).
First, we can see pentagons have smaller area on average ($\langle s_5 \rangle/\langle s \rangle<1$) and heptagons have larger area on average ($\langle s_7
\rangle/\langle s \rangle>1$). The trends of $\langle s_{5} \rangle/\langle s
\rangle$, $\langle s_{6} \rangle/\langle s
\rangle$ and $\langle s_{7} \rangle/\langle s
\rangle$ are more interesting.
Before jamming the trends depend on the initial packing
configuration of individual experiment, similar to the behavior of
$P(\bar{s})$ before jamming. However, close to and after the jamming
transition their features are robust for all experiments: $\langle
s_6 \rangle/\langle s \rangle$ plateaus, but $\langle
s_5 \rangle/\langle s \rangle$ clearly decreases and $\langle s_7
\rangle/\langle s \rangle$ increases. This is consistent with the behavior of glass-forming liquids \cite{Aharonov, Hentschel}.

Before we can explore the relation between the glass transition and the jamming transition further, we need to ask a simple but important question: where does an athermal system jam? Though the jamming point of a pack can be easily identified by its mechanical properties \cite{O'Hern}, experimentally it is usually hard to acquire further information of a pack except the position of all its particles. Can we detect the jamming transition of a granular pack from its structure? Despite the interesting features shown above, the Voronoi tessellation does not provide a unambiguous standard of the jamming transition. Both the area distribution and the annealing of defects show only quantitative changes before and after the transition. We shall show next that, for a pack of granular particles with purely repulsive interactions, its pair correlation provides a structural signature of the jamming transition.

\begin{figure}
\begin{center}
\includegraphics[width=3.35in]{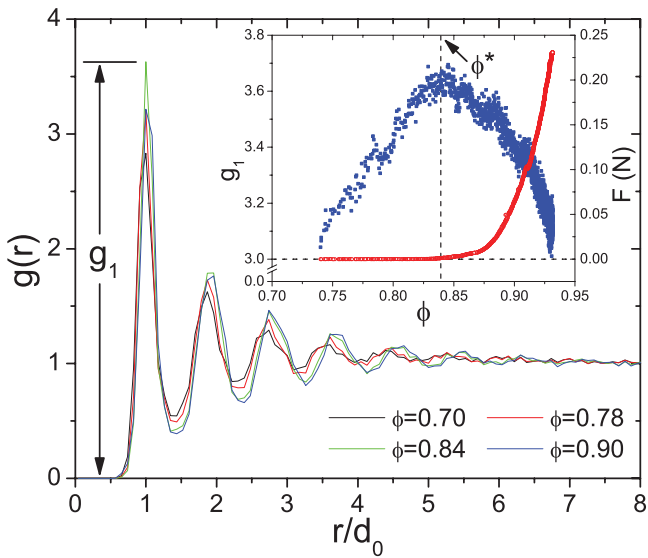}
\end{center}
\caption{Pair correlations of a pack at different
$\phi$. $r$ is normalized by the original size of pearls $d_0$. The
height of the first peak, $g_1$, is indicated. Inset: The height of the first peak $g_1(\phi)$ (blue squares) and the force measured at the
boundary of the cell $F(\phi)$ (red circles). The
vertical dashed line indicates the peak of $g_1(\phi)$, $\phi^*$. The
horizontal dashed line indicates the zero force.} \label{Figure3}
\end{figure}

We plot the pair correlation functions, $g(r)$, of a pack at
different $\phi$ through the jamming transition (Fig.~\ref{Figure3}). At each $\phi$, $g(r)$ has an oscillating
shape, typical for amorphous packs of spherical particles. Peaks in $g(r)$ indicate the layering structure in
the pack, the positions of which shift to larger $r$ as $\phi$
increases due to the enlargement of particles. A structural
signature of the jamming transition can be obtained by measuring $g_1$, the
height of the first peak of $g(r)$. $g_1(\phi)$ shows a non-monotonic behavior with
a peak at $\phi^*=0.84\pm0.02$ (Fig.~\ref{Figure3} inset). We also measured
the force applied by the pack on the boundary of the cell (Fig.~\ref{Figure3} inset). The
jamming point of the system is marked by the emergence of a
non-zero force on the boundary, which is coincident with the peak of
$g_1(\phi)$. Hence, $g_1(\phi)$ provides a clear structural
signature of the jamming transition and $\phi^*$ is the jamming
point.

The non-monotonic behavior of $g_1(\phi)$ can be seen as the vestige
of the structural singularity uncovered in the jamming transition of
monodisperse frictionless particles \cite{Silbert2}. As a system
evolves from unjammed state toward $\phi^*$, particles are pushed
closer to each other. The distribution of the distance between
neighboring particles becomes narrower and $g_1$ therefore
increases. For monodisperse frictionless particles, due to the
finite range purely repulsive interaction, at the jamming point the
distances between contact particles are all exactly one particle
diameter. Hence, $g_1$, representing the probability to find a
neighboring particle at a fixed distance, diverges. However, it
should be noted that for a system consisting of particles with
attractive interactions the onset of complete connectivity of all
particles and the rigidity of the pack need not occur at the same
packing fraction. Above $\phi^*$, due to the disordered nature of
the pack, particles will experience different compression and deform
differently depending on their local environment. The distances
between contact particles spread out again. Quantitatively,
$g_1\cdot w \sim z$ above jamming, where $w$ is the left-hand width
of the first peak of $g(r)$ and $z$ is the average coordination
number of particles \cite{Silbert2}. Since $z$ increases slowly, it
can be treated approximately as a constant near $\phi^*$. Hence, as
$w$ increases from zero above $\phi^*$, $g_1$ decreases. Apparently,
the polydispersity of our particles depresses the divergence of
$g_1$. Even at $\phi^*$, the distribution of the distances between
contact particles still has a finite width. Nevertheless, a
non-monotonic peak is preserved as a vestige of the singularity. A
simple estimation can be made on $g_1(\phi^*)$ with polydisperse
particles. Due to the difficulty to accurately measure the size of
swelled soft particles, we estimate the polydispersity of swelled
particles at $\phi^*$ to be around 5 to 8 percent, which is slightly
smaller than that of dry particles. If we assume that there is no
correlation between the radius of one particle and the radius of its
neighbors, $w$ should be roughly equal to the polydispersity of
particles, {\it i.e.} $w = 5\% \sim 8\%$ at $\phi^*$. Using the relation
$g_1(w)$ obtained from the simulations (Fig.2 of
Ref.\cite{Silbert2}), we found that $2.8<g_1(\phi^*)<4.5$, which is
consistent with our measurement (Fig.\ref{Figure3} inset). A similar
thermal vestige of the structural singularity has also been observed
in a colloidal system of finite $T$ \cite{Zhang}, where the
structural singularity is depressed by the thermal motion instead.

By analyzing the Voronoi tessellation and the pair correlation functions of a granular pack with increasing $\phi$, we have investigated the packing structures of the jamming transition. We have shown that the local configurations of a jamming granular pack are similar to those of glass-forming liquids, thereby demonstrating the analogy between the jamming transition and the glass transition in the spirit of the jamming phase diagram. Further study of the local structure of other jamming systems, such as glassy colloidal suspensions or jammed foams, may shed more light on the universal properties of the jamming structure illustrated here. Moreover, we have shown that a structural signature of the jamming transition can be obtained from the first peak of pair correlation functions. Our study shows that, despite the presence of realistic conditions such as frictional contacts and non-perfect spherical particle shapes, the structural signature observed in the ideal jamming system still persists in actual experiments.

The author thanks S. Nagel and H. Jaeger for their support and guidance. I am grateful to E. Brown, N. Keim, J. Royer, N. Xu and L.-N. Zou for their help with the experiments and fruitful discussions. I also thank L. Ristroph and S. Gerbode for comments on the manuscript. This work was supported by the NSF MRSEC program under DMR-0820054, by the Keck Initiative for Ultrafast Imaging at the University of Chicago and by the DOE under DE-FG02-03ER46088.

\end{document}